\documentclass[10pt,journal,final]{IEEEtran}

\ifCLASSINFOpdf
\usepackage[pdftex]{graphicx}
\else
\usepackage[dvips]{graphicx}
\fi
\usepackage[caption=false,font=normalsize,labelfont=sf,textfont=sf]{subfig}
\usepackage{amsfonts}
\usepackage{bm}
\usepackage{graphicx}	
\usepackage{multicol}
\usepackage{multirow}
\usepackage{tabulary}
\usepackage{tabularx}
\usepackage{stfloats}
\usepackage{mathrsfs}
\usepackage{latexsym}
\usepackage{epsfig}
\usepackage{epstopdf}
\usepackage{array}
\usepackage{amsmath}
\usepackage{amssymb}
\usepackage{color, soul}
\usepackage{amsthm}
\usepackage{enumerate}
\usepackage{algpseudocode}
\usepackage{algorithm}
\usepackage{bm}
\usepackage{balance}
\usepackage[noadjust]{cite}
\usepackage{float}
\usepackage{cite}
\usepackage{graphicx}
\usepackage{colortbl}
\usepackage{flushend}
\usepackage{graphics}
\usepackage{url}
\usepackage{lettrine}
\usepackage[utf8]{inputenc}
\usepackage{cite}
\usepackage[stable]{footmisc}

\usepackage{footnote}
\makesavenoteenv{center}
\makesavenoteenv{figure}  
\makesavenoteenv{table}   

\UseRawInputEncoding
\theoremstyle{plain} 

\theoremstyle{definition}

\def\bal#1\eal{\begin{align}#1\end{align}}

\newcommand{\bp} {\begin{proof}}
	\newcommand{\ep} {\end{proof}}

\newcommand{{\Rb}} {\right)}

\newcommand{{\Rf}} {\right\}}
\begin{document}
	
	\title{Deep CSI Feedback for FDD Massive MIMO Systems: A Curvelet Learning Approach} 

	\author{Mengli Tao,  Jiancun Fan, \emph{Senior Member, IEEE}, Huiqiang Xie, \emph{Member, IEEE}, Kai Xie
		\vspace*{-0.1in}
			\thanks{
				This work was supported by the National Natural Science Foundation of China under Grants No. 61671367 and No. 62401227. (\textit{Corresponding author: Jiancun Fan.})
				M. Tao and J. Fan are with the School of Information and Communication Engineering, Xi'an Jiaotong University, Xi'an, Shanxi 710049, P. R. China (e-mail: taoml@stu.xjtu.edu.cn; fanjc0114@gmail.com)
				
				H. Xie is with the School of Information Science and Technology, Jinan University, Guangzhou, Guangdong 510632, P. R. China (e-mail: huiqiangxie@jnu.edu.cn)

                Kai Xie is with Laboratory of Electromagnetic Space Cognition and Intelligent Control Technology, Beijing 100191, China (e-mail: eeixiekai@163.com)
				}
			}


	\maketitle
		\begin{abstract}
		 
       Downlink channel state information (CSI) feedback plays a key role in frequency division duplex (FDD) massive multiple-input multiple-output (mMIMO) systems. The growth of antennas in ultra-massive MIMO increases the difficulty and overhead of CSI feedback, which poses significant challenges for conventional downlink CSI feedback mechanisms. To address the limitations of existing CSI feedback approaches, this paper proposes a novel curvelet learning based framework termed SwinCANet, comprising a frequency-domain information processing module and a denoising module. The frequency-domain information processing module employs curvelet transform to decompose CSI into low-frequency and high-frequency components. Subsequently, Swin Transformer and channel-wise attention block are utilized for extracting the low-frequency and high-frequency representations, respectively, thereby enhancing reconstruction quality. Notably, an additional Swin Transformer facilitates the fusion of multi-scale frequency components, enhancing capabilities across different angular resolutions and spatial directions. Furthermore, we develop a variant (De-SwinCANet), which employs a Sigmoid threshold function to effectively suppress noise coefficients, thereby mitigating various channel impairments and nonlinear distortions. Numerical simulation results demonstrate that the proposed methodology achieves superior performance compared to existing benchmarks while maintaining robust performance under challenging propagation conditions.
		\end{abstract}
		\section{Introduction}
		

    \lettrine[lines=2]{M}{assive} Multiple-Input Multiple-Output (mMIMO) represents a pivotal technology in wireless communication systems, where both the base station (BS) and user equipment (UE) are equipped with substantially more antenna elements compared to conventional MIMO systems. Through the deployment of large-scale antenna arrays at both ends, mMIMO effectively enhances throughput, extends coverage range, and ensures the reliability of networks. This architectural advancement yields significant improvements in system capacity and spectral efficiency \cite{art_36}, making it indispensable for meeting the stringent performance requirements of future applications. Nevertheless, mMIMO systems encounter a fundamental challenge in practical deployment: optimal transmission performance is contingent upon the availability of highly accurate channel state information (CSI). In Frequency Division Duplex (FDD) systems, which are widely adopted in 5G networks, the principle of channel reciprocity becomes inapplicable due to the utilization of distinct frequency bands for uplink and downlink transmissions. Consequently, the BS cannot directly acquire downlink CSI and must rely on explicit feedback mechanisms from UE. The exponential increase in antenna array dimensions imposes substantial feedback overhead requirements that scale proportionally with the number of antenna elements, and aggravates the proportion of uplink resource consumption in FDD mode. Therefore, it is essential to develop efficient CSI feedback solutions to reduce feedback overhead and create effective techniques for compressing CSI matrices.
    

    Recently, deep learning (DL) has been widely adopted for channel feedback in massive MIMO systems \cite{art_3, art_7, art_8, art_9, art_10, art_11}, mitigating the stringent sparsity assumptions of conventional compressed sensing (CS). Research efforts have focused on convolutional neural networks (CNNs) and Transformers. Among CNN-based approaches, CsiNet \cite{art_3} pioneered DL-based CSI feedback by employing an encoder at the UE for compression and a decoder at the BS for reconstruction, outperforming CS methods. Subsequent enhancements include CsiNet+ \cite{art_7}, which improves performance via refined convolutional kernels; More recently, CRNet \cite{art_8} extracts multi-resolution features and highlights the importance of tailored training strategies for CSI feedback tasks. Ji \textit{et al.} \cite{art_9} introduced CLNet, which processes CSI in a physically meaningful complex-valued form via a pseudo-complex input layer and incorporates spatial attention to enhance performance. In \cite{art_10}, they further proposed a deep joint source-channel coding scheme employing nonlinear transformations for CSI compression, demonstrating advantages over separate source-channel coding. Wan \textit{et al.} \cite{art_11} developed a plug-and-play approach to reduce training and storage overhead in CSI feedback. Despite these advances, CNN-based networks remain limited in capturing global dependencies—a critical drawback in massive MIMO systems where long-range correlations across antenna pairs are prevalent. Due to constrained kernel sizes and hierarchical structures, CNNs primarily capture local features, potentially compromising reconstruction accuracy and system performance. Motivated by Transformer's capacity for global information modeling, researchers have explored Transformer-based architectures for CSI feedback. CsiTransformer \cite{art_12} first applied Transformer \cite{art_13} to this task, achieving marginal gains over CsiNet. Subsequently, TransNet \cite{art_14} adopted a dual-layer Transformer design to improve feedback accuracy. More recently, SwinCFNet \cite{art_15}, a Swin Transformer-based autoencoder, was proposed to effectively extract long-range dependencies in CSI.

     However, in practical massive MIMO systems, the escalation of antenna arrays from dozens to hundreds induces geometric growth in CSI dimensionality and a corresponding surge in feedback overhead. The resultant high dimensionality and intricate spatial correlations impose significant adaptive limitations on conventional neural network approaches. These challenges manifest primarily in two aspects: First, the dimensionality catastrophe leads to quadratic parameter growth, rendering computational complexity intractable for practical implementation. Second, the extremely high-dimensional input space significantly impedes training convergence and increases susceptibility to local optima. Third, \textcolor{black}{existing architectures show limited robustness in learning the intrinsic structured characteristics of mMIMO channels when the number of transmit antennas grows substantially.} Consequently, the development of specialized neural network architectures tailored to the unique characteristics of large-scale antenna arrays represents a critical research imperative. Such architectures must enhance CSI reconstruction accuracy while providing robust theoretical foundations and technical enablers for the advancement of mMIMO technologies.

     To address noise arising from feedback distortions and delays in CSI feedback, Jiang \textit{et al.} \cite{art_17} investigated deep autoencoder (AE)-based mechanisms explicitly considering these non-ideal factors in FDD massive MIMO systems. Concurrently, substantial research has targeted noise originating from channel estimation \cite{art_18, art_19, art_20}. Sun \textit{et al.} \cite{art_18} introduced AnciNet, a deep neural network designed to extract noise-free features from noisy CSI samples. Rahman \textit{et al.} \cite{art_20} proposed ResCBDNet, a residual convolutional blind denoising network incorporating accurate channel estimation models. To mitigate noise effects within codewords, Alam \textit{et al.} \cite{art_22} developed DCsiNet, where the decoder jointly performs decompression and denoising to enhance overall feedback accuracy. \textcolor{black}{To mitigate quantization distortion introduced by digital feedback channels, several studies have explored optimization within the joint source-channel coding (JSCC) framework. Guo \textit{et al.} \cite{art_16} proposed a CNN-based multi-rate CSI feedback scheme that reduces quantization effects via network optimization. Chen \textit{et al.} \cite{art_42} designed a quantizer with an offset network to compensate for quantization errors, albeit requiring multi-stage training. DeepCMC \cite{art_43} integrates quantization and entropy coding into end-to-end optimization, though its performance saturates at moderate-to-low compression ratios. Ravula \textit{et al.} \cite{art_40} introduced an autoencoder with an entropy bottleneck layer, which simulates quantization via uniform noise during training and employs uniform scalar quantization with arithmetic coding during inference. This enables joint optimization of quantization, entropy coding, and reconstruction, achieving superior rate-distortion performance across diverse bit rates.} Despite the demonstrated excellence of DL-based approaches in denoising applications, their inherent lack of model explanation presents significant limitations. The absence of interpretability not only impedes comprehensive understanding of model behavior but also constrains opportunities for systematic model optimization and theoretical advancement. \textcolor{black}{Although end-to-end neural networks can theoretically learn arbitrary transform domains, for massive MIMO channels with complex spatial non-stationarity, the large number of antennas will introduce the curse of dimensionality and increase the learning complexity.}
     In this paper, a novel channel feedback approach based on the curvelet transform is proposed to address the above challenges. The method utilizes the curvelet transform to decompose the CSI into low-frequency (global information) and high-frequency (local information) components. Specifically, we propose De-SwinCANet, which synergistically captures both global dependency patterns and local spatial characteristics of channels. Meanwhile, we also design a threshold function to process the channel feedback results to suppress the noise and improve the accuracy of the channel feedback. Finally, the CSI of the feedback is reconstructed by the inverse curvelet transform. The main contributions can be summarized as follows:
	
	\begin{itemize}
		\item[$\bullet$] This paper proposes a channel feedback denoising network called De-SwinCANet. This network is the first to apply the curvelet transform to the feedback and denoising of channels. 

		\item[$\bullet$] In the CSI feedback, we employ the curvelet transform to decompose CSI into low-frequency and high-frequency components, utilizing Swin Transformer to extract global structural features and channel-wise attention modules to enhance the utilization of local features, respectively. This separate processing method can improve the efficiency and accuracy of information extraction. By fusing low-frequency and high-frequency information with additional Swin Transformer blocks, it not only enhances the expressiveness of high-frequency information, but also provides rich multi-scale and directional features.  
		
		\item[$\bullet$] With theoretical analysis of noise in frequency domain, we design a Sigmoid threshold function to smoothly suppress small coefficients while preserving large coefficients, achieving effective signal-noise separation for denoising purposes, thereby enhancing the model's interpretability.
        
        \item[$\bullet$] Experiments demonstrate that under all signal-to-noise ratio (SNR) and compression ratio (CR) conditions, our model outperforms existing deep learning-based CSI reconstruction networks and reconstruction networks with denoising capabilities in CSI reconstruction, and the proposed method exhibits greater robustness across various channel conditions.
		
	\end{itemize}

	The remainder of the paper is organized as follows. In Section II, the system model is introduced. Section III introduces the architecture of the proposed model and introduces the curvelet transform. Simulations are provided in Section IV, and conclusions are drawn in Section V.
	
    $Notation$: Throughout this paper, bold letters are used to denote vectors or matrices, and regular letters are used to denote variables. $\mathbb{Z}$, $\mathbb{R}$, and $\mathbb{C}$ denote the integer field, the real field, and the complex field, respectively. ${( \cdot )^T}$ and ${( \cdot )^H}$ denotes the transpose and conjugate transpose, respectively; and $w\sim C{\cal N}(0,\sigma _n^2)$ denotes that the variable $w$ follows the circularly complex Gaussian distribution with zero mean and covariance $\sigma _n^2$.

		\section{System Model Framework}
				

    This work considers the downlink single cell frequency division duplex (FDD) mMIMO communication system, \textcolor{black}{consisting of ${N_t}$ antennas at the BS and a single antenna at the MS.} The system employs orthogonal frequency division multiplexing (OFDM) with $N_c$ subcarriers. The CSI of the downlink can be expressed as,
    \begin{equation}
        {\tilde{\bf H}} = {[{{\tilde{\bf h}}_1},{{\tilde{\bf h}}_2},...,{{\tilde{\bf h}}_{{N_c}}}]^T},
    \end{equation}    
    where ${\tilde{\bf H}} \in \mathbb{C} {^{{N_c} \times {N_t}}}$, \textcolor{black}{and $\tilde{\mathbf{h}}_n \in \mathbb{C}^{N_t \times 1}$ denotes the channel vector at the $n$-th subcarrier.}

    
    To design the precoding vector, the BS needs to obtain accurate CSI. In FDD systems, the UE transmits ${\tilde{\bf H}}$ back to the BS via the feedback link. However, direct feedback of the channel matrix ${\tilde{\bf H}}$ incurs enormous feedback overhead. Specifically, ${\tilde{\bf H}}$ contains $N_cN_t$ elements, which is prohibitive for capacity-limited feedback links and may severely constrain system performance.

    To reduce the overhead of channel feedback, we exploit the sparsity of the channel matrix ${\tilde{\bf H}} $ in the angular delay domain. Specifically, we first transform ${\tilde{\bf H}} $ into the angular delay domain using a two-dimensional (2D) discrete Fourier transform (DFT). With this transform, the energy can be concentrated on a few coefficients, thus achieving effective compression of the channel information and reduction of the feedback overhead, and we can obtain,
    \begin{equation}
        {\bar{\bf H}}={{\bf{F}}_c}{\tilde{\bf H}}{\bf{F}}_t^H,
    \end{equation}	
   where ${{\bf{F}}_c}$ and ${{\bf{F}}_t}$ represent the DFT matrices with dimensions ${{{\tilde N}_c}} \times {{{\tilde N}_c}}$ and ${N_t} \times {N_t}$, respectively. For the angular-delay domain channel matrix ${{{\bar{\bf H}}}}$, only the first ${N_a}$ rows contain significant values, while the remaining rows consist of elements close to zero, which can be omitted with minimal information loss. For ease of understanding, we denote the first ${N_a}$ rows of ${{{\bar{\bf H}}}}$ as ${{\bf{H}}}$.
    
    Despite ${{\bf{H}}}$ being smaller than ${\bar{\bf H}}$, ${N_{a}}{N_{t}}$ remains large, requiring further compression. However, angular pruning makes ${\bf{H}}$ dense rather than sparse, limiting CS-based compression effectiveness. NN-based methods overcome this limitation by learning nonlinear channel characteristics and achieving superior reconstruction even under low sparsity, outperforming CS-based approaches in both accuracy and feedback performance.
    
    \begin{figure}[t]  
            \centering
            \includegraphics[width=8cm]{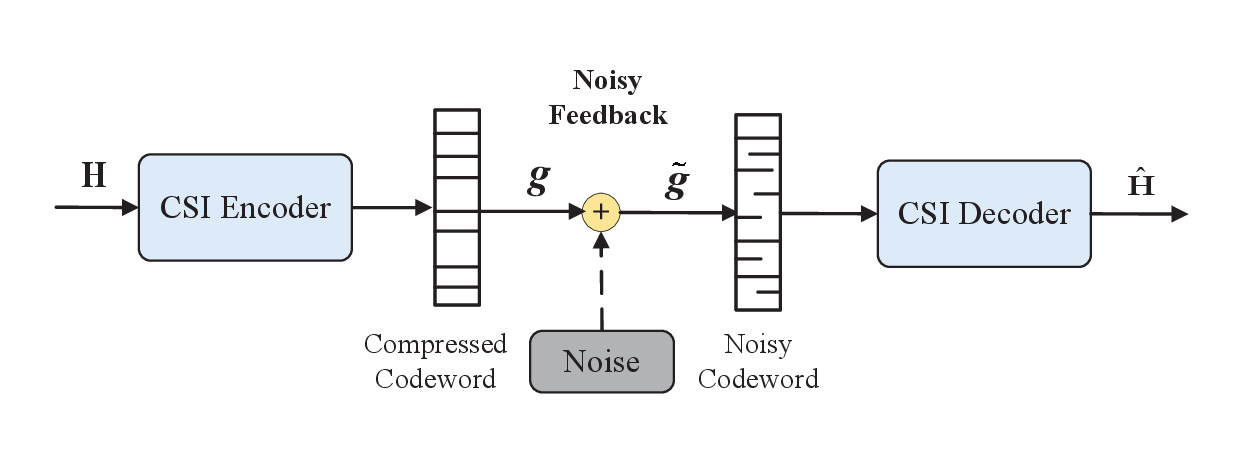}
            \captionsetup{justification=centering}
            \caption{\textcolor{black}{The Structure of the CSI feedback model.}}
            \label{Fig. 1.}           
    \end{figure}
    
    Fig. 1\ref{Fig. 1.} illustrates the feedback process of CSI. First, the UE side extracts the key features of ${\bf{H}}$ through the encoder and outputs the compressed codeword as follows:
    \begin{equation}
        {\bf{g}} = {{\cal F}_\varepsilon }({\bf{H}};{\Theta _\varepsilon }),
    \end{equation}
    where ${\bf{g}} \in\mathbb{R} {^M}$ denotes the compressed codeword, and ${{\cal F}_\varepsilon }$ and $\Theta _\varepsilon$ denote the coding process and a set of parameters of the encoder, respectively. When $M < 2{N_a}{N_t}$, the compression ratio $\gamma = M/2{N_a}{N_t}$.

    \textcolor{black}{We adopt the JSCC framework for CSI feedback due to its lower quantization errors and greater robustness, in which the encoder's codewords are transmitted directly over the wireless channel.} The codeword $\mathbf{g}$ output by the encoder is transmitted directly through the wireless channel. The channel effect can be modeled as additive white Gaussian noise (AWGN) with fading characteristics \cite{art_28}, which is treated as an additional component superimposed on the codeword,


\begin{equation}
    {\mathbf{y}} = \left\{ \begin{array}{ll}
        \mathbf{g} + \mathbf{n}, & \text{AWGN channel}, \\
        h\mathbf{g} + \mathbf{n}, & \text{slow fading channel},
    \end{array} \right.
\end{equation}
    where $h \in \mathbb{C}$ denotes the channel gain and $\mathbf{n} \in \mathbb{C}$ represents the AWGN term. \textcolor{black}{In practical systems, the receiver performs channel estimation using pilot signals. Considering the presence of estimation errors, let the estimated channel gain be $\hat{h} = h + \Delta h$, where $\Delta h \sim \mathcal{CN}(0, \sigma_e^2)$ denotes the channel estimation error, with $\sigma_e^2$ depending on the pilot power, the number of pilots, and the channel estimation algorithm. Using the estimated channel gain $\hat{h}$ for equalization, the receiver obtains $\tilde{{\mathbf g}} = \frac{{\mathbf{y}}}{\hat{h}} = \frac{h}{h + \Delta h}\mathbf{y} + \frac{\mathbf{n}}{h + \Delta h}$.}

    After receiving the noisy codeword ${\tilde{\bf g}}$, the UE reconstructs the channel matrix using the decoder, that is,
     \begin{equation}
        {\hat { \bf{H}}} = {{\cal F}_\chi }({\tilde{\bf g}};{\Theta _\chi }),
    \end{equation}
    where ${\hat {\bf{H}}}$ denotes the reconstructed codeword, ${{\cal F}_\chi }$ and $\Theta _\chi$ denote the decoding process and a set of parameters of the decoder, respectively.

    \textcolor{black}{Existing CS-based channel feedback methods suffer from two fundamental limitations. First, their performance relies heavily on the channel being strictly sparse, and significant degradation may occur in practical scenarios where the channel exhibits non-strict sparsity \cite{art_21}. Second, these methods lack robustness, as the reconstruction algorithms are highly sensitive to noise and interference in the feedback link, making accurate channel recovery difficult in non-zero-noise environments \cite{art_20}.} Although DL-based denoising methods can achieve certain effectiveness, they exhibit high computational complexity and lack solid theoretical foundation. Therefore, this paper proposes a curvelet transform-based deep CSI feedback network.

    \section{Design of De-SwinCANet}
     
        In this section, we first describe the framework of the proposed De-SwinCANet. The proposed De-SwinCANet consists of the SwinCANet for channel feedback and the threshold function for denoising. Our introduction follows this order: First, we elaborate on the theoretical foundation of the curvelet transform; next, we describe the network architecture of SwinCANet in detail; finally, we introduce the Sigmoid threshold function used to implement the denoising function.

        \begin{figure*}[t]  
            \centering
            \includegraphics[width=16.5cm]{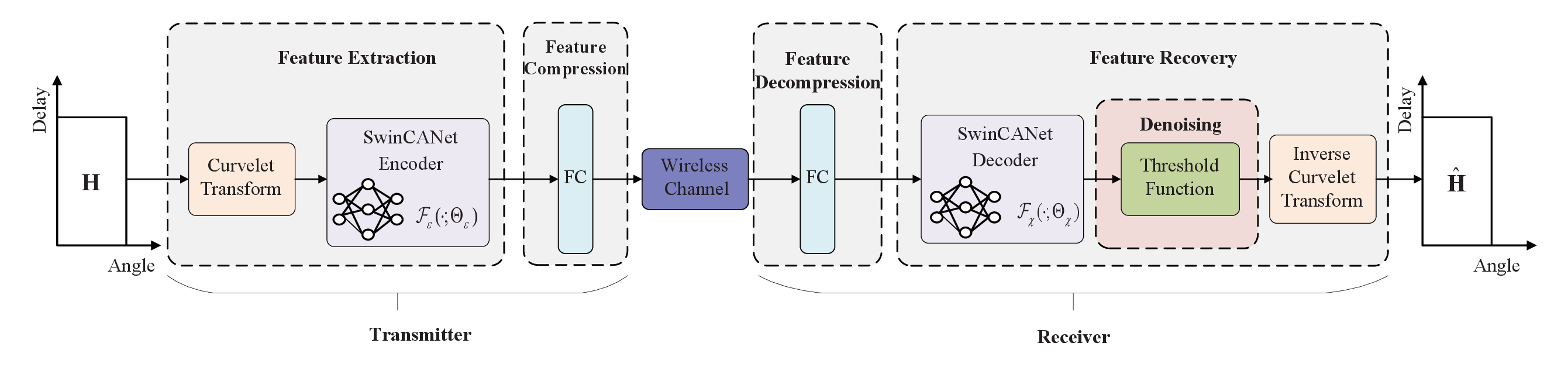}
            \captionsetup{justification=centering}
            \caption{The proposed De-SwinCANet model.}
            \label{Fig. 2-1.}           
        \end{figure*}
        
    \subsection{Structure of De-SwinCANet}

        \emph{1) Overview and Settings}: Fig. \ref{Fig. 2-1.} shows the framework of the proposed De-SwinCANet. Specifically, the encoder consists of two parts: the feature extraction part and the feature compression part. The feature extraction part is implemented by curvelet transform and SwinCANet encoder. \textcolor{black}{The input CSI matrix $ \bf{H} $ of size $2 \times N_a \times N_t $, obtained via DFT, subsequently undergoes frequency domain information decomposition through the curvelet transform (thus the curvelet transform is applied after the DFT in this work), yielding the curvelet coefficient matrix.} Next, the curvelet coefficient matrix is processed through the SwinCANet encoder. The feature compression stage reduces feature dimensions through fully connected (FC) layers. The compressed signal is subsequently transmitted through the noisy channel.

        The decoder is divided into two stages: feature decompression and feature recovery. The feature decompression stage uses FC layers to realize feature upscaling, while the feature recovery stage consists of the SwinCANet decoder, denoising module, and inverse curvelet transform. The SwinCANet decoder first decodes the noisy codeword $ {\tilde{\bf g}} $ to recover the original features. Then, the denoising module filters the obtained curvelet coefficient matrix to remove small coefficients and suppress noise. Finally, the denoised CSI matrix is reconstructed through the inverse curvelet transform.

        \begin{figure}[t]  
            \centering
            \includegraphics[width=8cm]{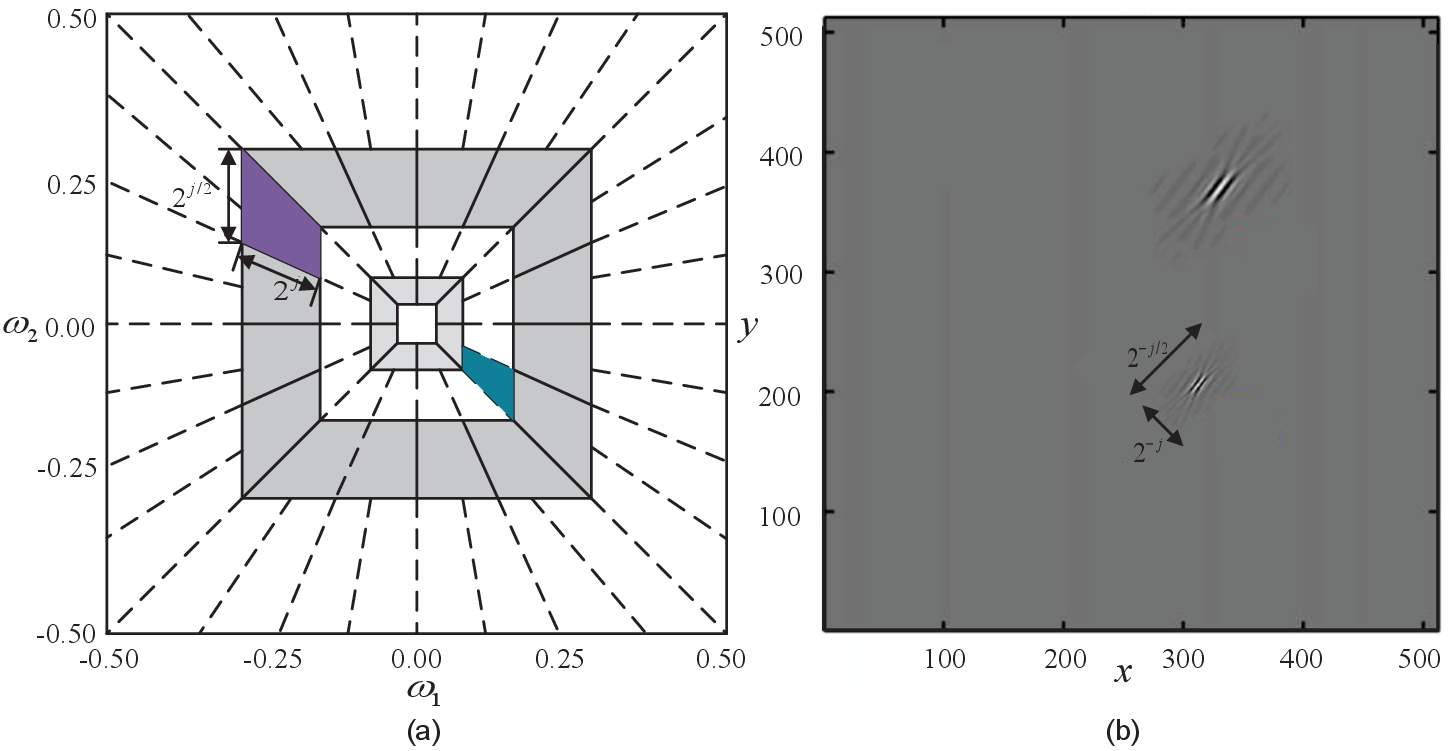}
            \captionsetup{justification=centering}
            \caption{(a) Wedges in wavenumber domain; (b) The spatial domain curvelets correspond to wedges in the wavenumber domain.}
        \end{figure} 

        \begin{figure*}[t]  
            
            \centering
            \includegraphics[width=16cm]{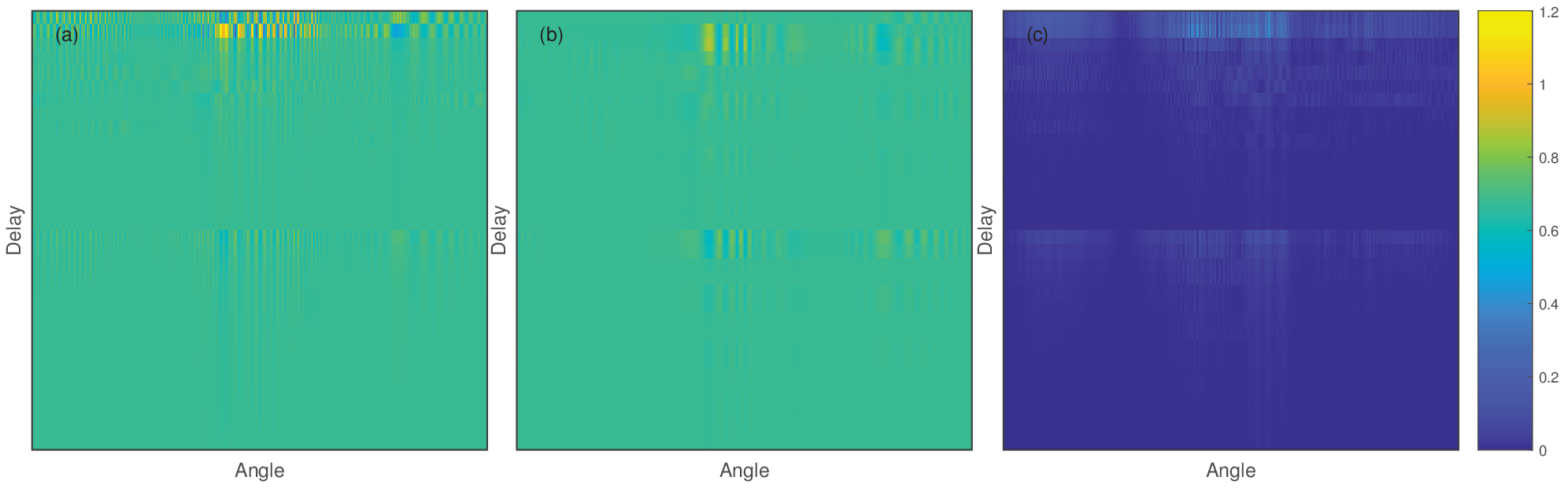}
            \captionsetup{justification=centering}
            \caption{CSI and its reconstruction results at different scales (a) Original signal; (b) First-scale (low-frequency) reconstruction results; (c) Second-scale (high-frequency) reconstruction results.}
           
        \end{figure*}

    \subsection{Curvelet Transform}


        \textcolor{black}{The curvelet transform \cite{art_23}, proposed by Candes and Donoho in 1999, is a multi-scale, multi-directional mathematical transform that provides near-optimal sparse representations for signals with curve-like singularities. Compared to wavelet transforms, curvelets offer superior directional selectivity, making them particularly effective for representing edge-like features in 2D signals. To our knowledge, our work is the first to apply the curvelet transform to CSI feedback in wireless communications.}

        \textcolor{black}{We employ the discrete curvelet transform based on the wrapping technique \cite{art_24} to decompose the CSI matrix $\mathbf{H}$ into multi-scale components. The transform first performs a 2D FFT on $\mathbf{H}$ to obtain the Fourier samples $\mathbf{H}'$. For each scale and angle pair $(j, l_j)$, the Cartesian window $\tilde{U}_{j,l_j}[n_a, n_t]$ is multiplied with the Fourier samples to obtain:}
        \begin{equation}
            \textcolor{black}{Q{j,l_j}[n_a, n_t] = \tilde{U}_{j,l_j}[n_a, n_t]\mathbf{H}'[n_a, n_t].}
        \end{equation}
        \textcolor{black}{Then, the wrapping operation maps the wedge-shaped regions to rectangular regions around the origin:} 
        \begin{equation}
            \textcolor{black}{\mathbf{H}'_{j,l_j}[n_a, n_t] = \text{Wrapping}(\mathbf{H}'[n_a, n_t]) = W(Q_{j,l_j})[n_a, n_t].}
        \end{equation}
        \textcolor{black}{As shown in Fig. 3(b), this step maps information from a specific scale and angle (purple region) to a rectangular region around the origin. Due to the 2D Fourier transform's periodicity, this mapping preserves all information without loss.}
        
        \textcolor{black}{After applying 2D-IFFT on each wrapped frequency domain data, we obtain the complete discrete curvelet coefficient set:}
        \begin{equation}
            \textcolor{black}{\mathcal{C} = \text{Curvelet}_J(\mathbf{H}') = \left\{ \bm{\mathcal{C}}_0, \left\{ \bm{\mathcal{C}}_j \right\}_{j=1}^{J-1} \right\},}  
        \end{equation}
        \textcolor{black}{where $\mathcal{C}_0 \in \mathbb{R}^{C_a \times C_b \times 2}$ represents the coarse-scale coefficients capturing low-frequency components, and $\{{\bm{\mathcal{C}}}_j\}_{j=1}^{J-1}$ with $\mathcal{C}_j \in \mathbb{R}^{S_{aj} \times S_{bj} \times 2}$ represents the fine-scale coefficients containing high-frequency details. The inverse transform reconstructs the signal as: }
        \begin{equation}
            \textcolor{black}{{\hat {\bf H}} = {\rm{Inv\_Curvele}}{{\rm{t}}_J}(\bm{\mathcal{C}}).}
        \end{equation}
        \textcolor{black}{The specific implementation steps are summarized in Algorithm 1.}
        
\begin{algorithm}[h]
    \caption{The process of the curvelet transform.}
    \label{2}
    {\bf Input:} 
    For the CSI $\bf{H}$ in the angle-delay domain to be fed back, set up a $J$-level curvelet transform.  
        \begin{algorithmic}[1]
            \State \! Obtain the discrete Fourier sampling $\bf{H'}$ of the CSI matrix through 2D FFT.         
            \Statex Take the Cartesian window and multiply it with the Fourier samples:
            \For{ each $(j,l_j)$}:         
                    \State \! obtain $Q_{j,l_j}$ through eq (6).
            \EndFor
            
            \Statex Periodically map the wedge regions to rectangular regions:
            \For{ each $(j,l_j)$}:
                \State \! Obtain the coefficients of $Q_{j,l_j}$ wrapped around the origin of the Cartesian coordinate system through eq (7).
            \EndFor   
            
            \Statex Apply 2D-IFFT on each wrapped frequency domain data:
            \For{ each $(j,l_j)$}:
                \State \! Obtain the discrete curvelet coefficients through eq (8).
            \EndFor

        \end{algorithmic}
        
    {\bf Output:} 
    Complete discrete curvelet coefficients: ${\bm{\mathcal{C}}}=\{{\bm{\mathcal{C}}}_0,\{{\bm{\mathcal{C}}}_j\}_{j=1}^{J-1} \}$
    
\end{algorithm}

         As shown in Fig. 3(b), this step maps information from a specific scale and angle (purple region in upper left) to a rectangular region around the origin. The wrapping technique translates all frequency domain regions to a rectangular affine region at the origin. Due to the 2D Fourier transform's periodicity, this mapping preserves all information without loss. Fig. 3(b) demonstrates curvelet spatial domain characteristics. When the spatial scale is $2^{-j}$ (Fig. 3(a)), the corresponding curvelet has support length $2^{-j/2}$ and width $2^{-j}$. The parabolic relationship $width \approx length^2$ reveals curvelets' significant directional and anisotropic properties. Curvelet coefficients are organized as unit arrays with scale- and orientation-specific matrices of varying sizes. Transform implementation involves inverse transforms on beam domain signals: the beam domain partition grids in Fig. 3(a) convert to spatial domain curvelets in Fig. 3(b), where purple and blue regions correspond to wedge window functions at the fourth and third scales, respectively. 
         


        To assess the curvelet transform's feature extraction and reconstruction efficacy in multi-scale signal analysis, this study employs scale-wise reconstruction for quantitative evaluation, emphasizing the impact of edge signal recovery across frequency scales. Utilizing the CSI matrix (Fig. 4(a)), a two-level curvelet decomposition is performed to derive multi-scale coefficient representations, followed by independent reconstruction of each scale to validate the transform's scale separation properties. The reconstruction outcomes (Fig. 4(b)-4(c)) illustrate the curvelet transform's proficient separation of multi-scale signal features. Coarse-scale reconstruction retains global contours and low-frequency information, demonstrating robust macroscopic structure characterization (Fig. 4(b)). In contrast, fine-scale reconstruction highlights local details and microscopic variations, sensitively capturing edge effects from boundary truncation discontinuities in the measurement domain (Fig. 4(c)). The experimental results validate the curvelet transform's capability for multi-scale decomposition, effectively separating global structures from local details (including truncation edge effects) across different coefficients. This scale-adaptive property establishes a theoretical foundation for refined analysis of complex signals.

        \subsection{Structure of SwinCANet}
        
        \emph{1) Design of the Architecture}: In the feature extraction stage, the low-frequency and high-frequency components of the CSI are effectively separated by the curvelet transform. We can obtain the coarse-scale coefficient matrix (characterizing the low-frequency component of the signal) ${\bm{\mathcal{C}}}_0 $ and the set of fine-scale directional subbands (characterizing the high-frequency component of the signal) $\{{\bm{\mathcal{C}}} _j\}_{j=1}^{J-1} \}$. Except for the coarse scale ${\bm{\mathcal{C}}}_0$, each scale contains multiple directional components, which represent the local structural information in different directions at that scale.

        Curvelet transform's low-frequency coefficients serve as Swin Transformer input, achieving multi-level collaborative optimization. \textcolor{black}{The curvelet transform decomposes the CSI into low-frequency and high-frequency components. Among these, the low-frequency coefficients capture the global structure and macro trends of the channel matrix \cite{art_24}, which naturally aligns with Swin Transformer's strength in modeling long-range dependencies \cite{art_32}. Moreover, the low-frequency coefficients preserve the primary energy of the signal while reducing dimensionality. Using these low-resolution, low-redundancy low-frequency components as input to the Swin Transformer, rather than the original high-dimensional CSI, can significantly alleviate the computational burden on the Swin Transformer. Furthermore, since signal energy is concentrated in the low-frequency band while noise energy is dispersed across the entire frequency band \cite{art_39}, the low-frequency components inherently possess a higher SNR, thereby providing a more robust feature representation for subsequent DL models.} This combined mathematical transform and deep learning approach provides an effective balance between accuracy and efficiency for low-frequency information processing. \textcolor{black}{Besides, high-frequency coefficients contain rich local information, which is crucial for signal reconstruction and detail enhancement. The high-frequency components exhibit strong sparsity, with most coefficients being zero or close to zero. This structured frequency-domain representation naturally supports a high compression ratio and significantly reduces the input dimensionality for subsequent neural processing. We introduce a channel-wise attention mechanism \cite{art_30} that adaptively learns the importance distribution of each channel and dynamically reweights the high-frequency coefficients. This mechanism effectively enhances the representation of key local features while suppressing noise and redundant components, thereby achieving a synergistic optimization of detail preservation and redundancy reduction.}

        \begin{figure*}[t]  
            \centering
            \includegraphics[width=15cm]{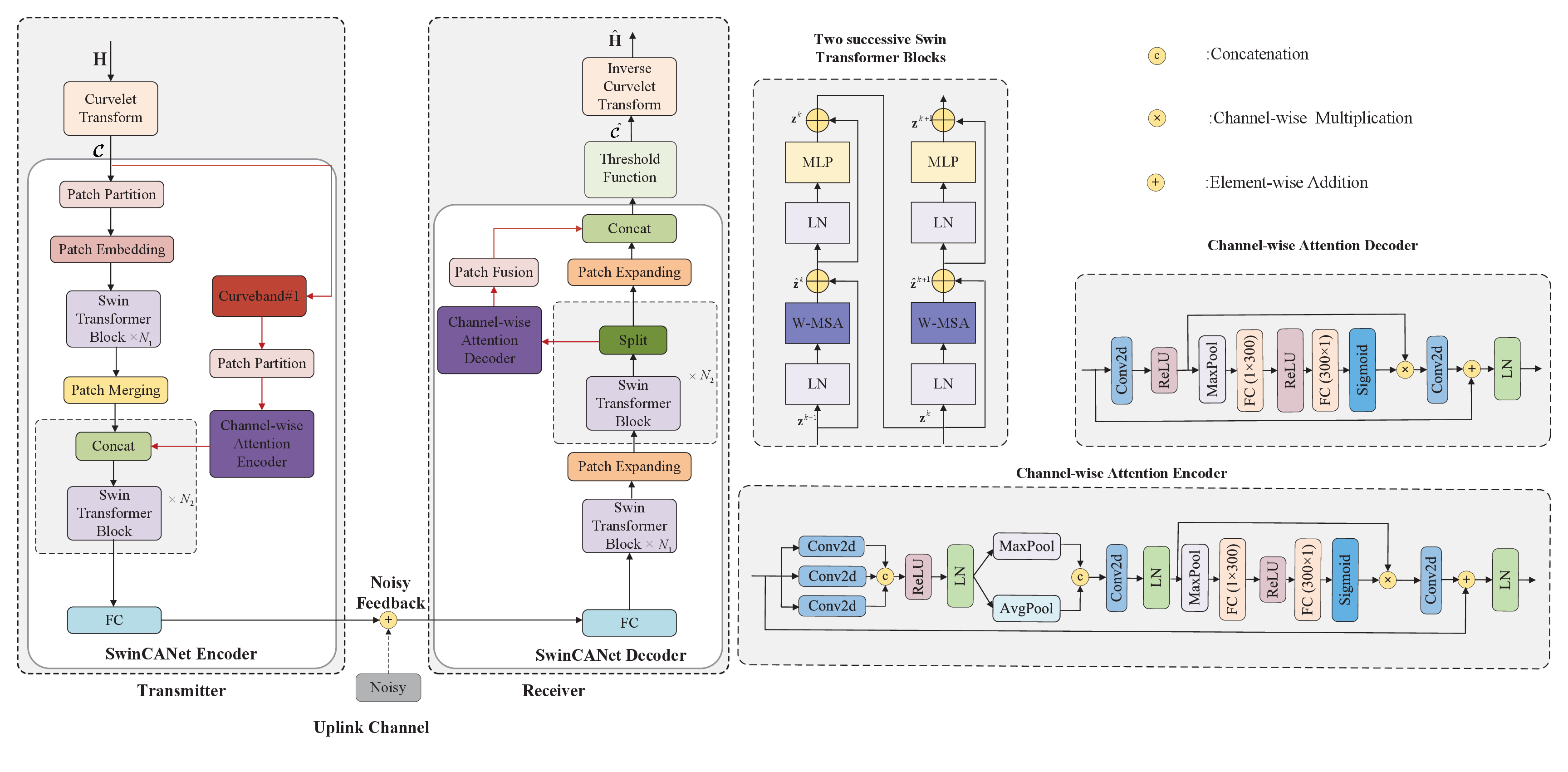}
            \captionsetup{justification=centering}
            \caption{The overall architecture of the proposed De-SwinCANet.}
        \end{figure*}         
       
        Next we describe the proposed SwinCANet in detail, as shown in Fig. 5, the system consists of an encoder on the UE and a decoder on the BS. Unlike models that only use Swin Transformer, we add a channel-wise attention block to the model for extracting higher-scale high-frequency information. The low-frequency information is then fused with high-frequency information in the Swin Transformer block in the second stage to enhance the feature representation. In this way, we are able to efficiently perform multi-scale learning and extract orientation information.
        
        \emph{2) Coarse-scale Feature Extraction with Swin Transformer}: The coarse-scale information in the curvelet domain is represented as a 3D input image of size ${C_a} \times {C_b} \times 2$, where the last dimension 2 represents the real and imaginary parts of the coarse-scale curvelet coefficients, respectively. In the encoder part, the coarse-scale curvelet coefficient matrix is processed by the patch partition module of ${\rm{3}} \times {\rm{3}}$, which splits the matrix into multiple patches and treats each patch as a token to realize the conversion to sequence embedding. The patch partition module divides the coarse-scale coefficient matrix into $\frac{{{C_a}}}{{\rm{3}}} \times \frac{{{C_b}}}{{\rm{3}}}$ non-overlapping patches. With this division method, the feature dimension of each patch becomes ${\rm{3}} \times {\rm{3}} \times 2$. Subsequently, the patch embedding layer projects the patch features into a predefined embedding dimension $E_1$, yielding ${\bf {z}}^0$. Then passes through $N_1$ Swin Transformer blocks to maintain the number of tokens.
        
        Unlike convenional Multi-Head Self-Attention (MSA) modules, the Swin Transformer block \cite{art_32} is constructed based on shifted windows. As shown in Fig. 5, two successive Swin Transformer blocks are illustrated, where each block consists of a Layer Normalization (LN) layer, a multi-head self-attention module, a residual connection, and a two-layer MLP with GELU activation. Specifically, a Window-based Multi-Head Self-Attention (W-MSA) module and a Shifted Window-based Multi-Head Self-Attention (SW-MSA) module are alternately applied in two successive Transformer blocks, enabling efficient local modeling while achieving cross-window information interaction through window shifting. Following this window partitioning approach, continuous Swin Transformer blocks can be designed as,
        \begin{equation}
            \begin{aligned}
                \hat{\bf{z}}^k &= \text{W-MSA}(\text{LN}({\bf{z}}^{k-1})) + {{\bf{z}}^{k-1}}, \\
                {\bf{z}}^k &= \text{MLP}(\text{LN}(\hat{\bf{z}}^k)) + \hat{\bf{z}}^k ,\\
                \hat{\bf{z}}^{k+1} &= \text{SW-MSA}(\text{LN}({\bf{z}}^k)) + {{\bf{z}}^k}, \\
                {\bf{z}}^{k+1} &= \text{MLP}(\text{LN}(\hat{\bf{z}}^{k+1})) + \hat{\bf{z}}^{k+1},
            \end{aligned}
        \end{equation}
        where $\hat{\bf{z}}^k$ and ${\bf{z}}^k$ denote the outputs of the (S)W-MSA module and the MLP module in the $k$-th block, respectively.

        The self-attention is computed as follows:
        \begin{equation}
             {f_{\rm SA}}({\bf z}^k) = {\rm{softmax}}(\frac{{{{\bm Q}}{\bm K}^T}}{{\sqrt {{d}} }}){{\bm V}},
         \end{equation}
        where ${{\bm Q}}={\bf z}^k {\bm W}^{Q} $, ${{\bm K}}= {\bf z}^k {\bm W}^{K}$, and ${{\bm V}}= {\bf z}^k {\bm W}^{V}$, in which ${\bm W}^{Q}, {\bm W}^{K}$, and ${\bm W}^{V}$ are the learnable weights across different windows. The operator ${\rm{softmax(}} \cdot {\rm{)}}$ represents the softmax activation function.

         After passing through $N_1$ Swin transformer blocks, then is fed into a patch merging layer to reduce the feature size.

        \emph{3) Fine-scale information extraction via channel-wise attention mechanism}: For the high-frequency components of CSI, the system adopts a multi-angle cyclic processing architecture for processing. The coefficient matrix at the $j$th scale is represented as a 3D input image of ${{S_{a_j}} \times {S_{b_j}} \times 2}$, where the last dimension 2 denotes the real and imaginary parts of the curvelet coefficients, respectively. After the reshaping operation is also first processed by the patch division module to realize the sequence embedding to obtain ${\bar{\bm{\mathcal{C}}}}_j \in \mathbb{R}^{{S'_{a_j}} \times {S'_{b_j}} \times 2}$, and then enters the channel-wise attention module processing, which performs adaptive reweighting of high-frequency coefficients. In this process, each channel corresponds to a specific feature map, and these feature maps are generated through learnable filters, representing different learned feature patterns respectively \cite{art_30, art_31}. Therefore, the purpose of the channel-wise attention mechanism is to highlight important high-frequency detail information. Next, we will introduce the detailed architecture of the channel-wise attention mechanism.
        
        On the encoder side, we first apply convolutional kernels with varying window sizes to the input ${\bar{\bm{\mathcal{C}}}}_j$ to capture multi-scale spatial features. The resulting feature maps are processed through both max pooling (MP) and average pooling (AP) operations to extract complementary spatial information, which are then concatenated and further refined using a convolutional block. A simple gating mechanism, implemented via two fully connected layers with a sigmoid activation in between, is used to generate channel attention weights. These attention weights are applied to the normalized features via channel-wise multiplication. The attended features are then transposed, and a final 2D convolutional layer is used to reconstruct the channel domain. Finally, a residual connection adds the original input ${\bar{\bm{\mathcal{C}}}}_j$ back to the processed features to obtain the final output of the channel-wise attention mechanism. And the channel-wise attention mechanism at the $j$-th subbands consists of the following components,
        \begin{equation}
        \begin{aligned}
            \begin{array}{l}
            {f_{CA}}({\bar{\bm{\mathcal{C}}}}_j) = {\rm{LN(}}{f^{{C_5}}}({\rm{sigmoid}}({{\bm{W}}_2}{{\bm{W}}_3}({\rm{MP}}({\rm{LN}}({f^{{C_4}}}(\\
            \begin{array}{*{20}{c}}
            {}&{}&{\begin{array}{*{20}{c}}
            {}&{}
            \end{array}}&{[{\rm{MP}}({\bar{\bm{\mathcal{C}}}}_j);{\rm{AP}}({\bar{\bm{\mathcal{C}}}}_j)]))))) \odot {{\tilde {\bm{\mathcal{C}}}}_j}{)^T} + {\bar{\bm{\mathcal{C}}}}_j)}
            \end{array}
            \end{array},
         \end{aligned}    
         \end{equation}
         where $f(\cdot)$ is the 2D convolutional layer, ${\tilde {\bm{\mathcal{C}}}}_j = {\delta}[{{\bm C}_1};{{\bm C}_2};{{\bm C}_3}]$, ${\bm C}_1=f^{C_1}({\bar{\bm{\mathcal{C}}}}_j)$, ${\bm C}_2=f^{C_2} ({\bar{\bm{\mathcal{C}}}}_j)$, and ${\bm C}_3=f^{C_3} ({\bar{\bm{\mathcal{C}}}}_j)$, where $f^{C_1}_i(\cdot)$, $f^{C_2}_i(\cdot)$, and $f^{C_3}_i(\cdot)$ are the 2D convolutional layers with different kernel sizes, and $\delta(\cdot)$ represent the ReLU activate function. ${\bm W}_2$ and ${\bm W}_3$ are two FC layers, representing a simple gating mechanism used to obtain the attention weights for each channel, capturing the dependencies between channels to effectively capture the key feature information of the CSI.

        The high-frequency information from the channel-wise attention module is concatenated with the preprocessed low-frequency information, forming inputs that integrate directional features and global context. To fully capture multi-angle directional characteristics, $N_2$ Swin Transformer blocks are cascaded, corresponding to the number of angles. Following the layer-by-layer progressive mechanism of Swin Transformer, features from each angle are iteratively processed by the corresponding blocks, where the output of each stage serves as the input for the next. This recurrent refinement enables mutual interaction and progressive fusion of directional features across different angles. Each processing step performs feature refinement and enhancement based on the previous step, ultimately forming a comprehensive feature representation containing rich directional information, which provides a high-quality feature foundation for subsequent curvelet coefficient reconstruction and signal recovery. Finally, a fully connected layer is used to obtain the compressed codeword $\bf{g}$.
        
        After passing through the wireless channel, we obtain the noise signal ${\tilde{\bf{g}}}$. On the decoding side, the SwinCANet decoder ${{\cal F}_\chi }( \cdot ;{\Theta _\chi })$ is designed as a symmetric part with the encoder ${{\cal F}_\varepsilon }( \cdot ;{\Theta _\varepsilon })$. First, decompression is performed through an FC layer. Feature recovery comprises patch expanding layers, Swin Transformer blocks, and channel-wise attention decoding modules. The decoding process uses pre-specified Swin Transformer blocks and patch expanding layers for upsampling to recover compressed features to appropriate scales. The system employs cyclic decoding architecture to progressively reconstruct curvelet coefficients for each angle. Data feeds into angle-corresponding Swin Transformer blocks, processed cyclically following standard Swin Transformer decoding procedures. After each Swin Transformer block, data splits into low-frequency and high-frequency components. Low-frequency information passes to the next block as basic information for subsequent angle processing. Through cyclic decoding across all blocks, the system gradually separates and reconstructs coefficient matrices for all angles, achieving step-by-step decomposition from comprehensive features to specific angle coefficients. High-frequency information ${\hat {\cal C}}_j$ enters the channel-wise attention decoding module for feature recovery, where attention mechanisms restore directional detail features for each angle. The corresponding decoding process can be described by the following formula,
        \begin{equation}
             \begin{array}{l}
            {f_{CA}}({\hat {\bm{\mathcal{C}}}}_j) = {\rm{LN(}}{f^{{C_7}}}({\rm{sigmoid}}({{\bm{W}}_4}{{\bm{W}}_5}({\rm{MP}}(\delta (f_{^i}^{{C_6}}({\hat {\bm{\mathcal{C}}}}_j))))) \cdot \\
            \begin{array}{*{20}{c}}
            {}&{}&{}&{\begin{array}{*{20}{c}}
            {}&{}
            \end{array}}
            \end{array}{{\tilde {\hat {\bm{\mathcal{C}}}}}_j}{)^T} + {\hat {\bm{\mathcal{C}}}}_j),
            \end{array}
         \end{equation}
        where ${{\tilde {\hat {\bm{\mathcal{C}}}}}_j}={\delta}({{\bm C}_5})$, ${{\bm C}_5}=f^{C_5} ({\hat {\bm{\mathcal{C}}}}_j)$, $f^{C_6}_i(\cdot)$ and $f^{C_7}_i(\cdot)$ are the 2D convolutional layers, and ${\bm W}_4$ and ${\bm W}_5$ denote the FC layers, respectively.
        
        Finally, through patch fusion modules and reshaping operations, the reconstructed high-frequency curvelet coefficients are obtained. Meanwhile, the ultimately separated low-frequency information is processed by patch expanding layers for upsampling to obtain reconstructed low-frequency coefficients. Eventually, the coefficient matrices from all angles, together with the separated low-frequency information, constitute a complete curvelet coefficient representation, providing an accurate coefficient foundation for subsequent inverse curvelet transform and signal reconstruction.

        \begin{figure}[t]  
            
            \centering
            \includegraphics[width=8cm]{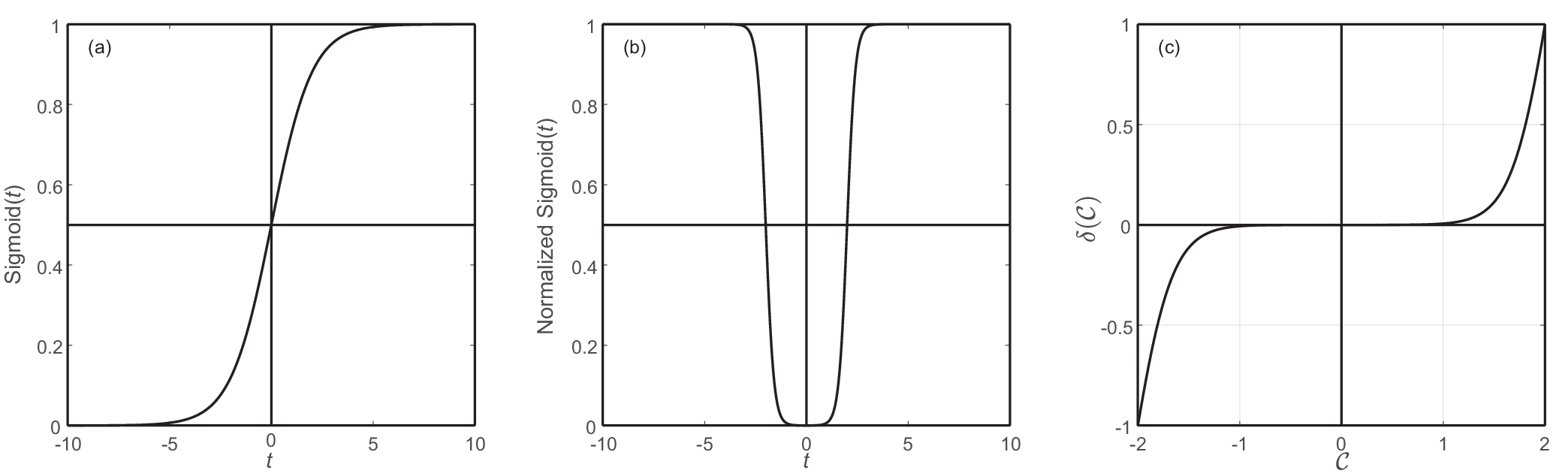}
            \captionsetup{justification=centering}
            \caption{(a) Sigmoid function; (b) Normalized Sigmoid function ($T$ = 2); (c) Sigmoid threshold function ($T$ = 2).}
           
        \end{figure}

        \subsection{The Setting of the Threshold Function}
         According to Candes and Donoho \cite{art_25}, when the signal features match the curvelet basis functions well, the corresponding curvelet coefficients are large; conversely, when the match is poor, the coefficients are small. Since the major components of a useful signal typically exhibit regular and smooth variations, they can be described by larger curvelet coefficients. In contrast, noise often exhibits random oscillatory characteristics and is described by smaller curvelet coefficients. Based on this property, we set thresholds at each scale in the curvelet domain to filter out coefficients related to noise, and then reconstruct the remaining curvelet coefficients back into the spatial domain for denoising.

        Based on the above analysis, we know that the choice of threshold function is crucial, as it determines whether the curvelet coefficients of the useful signal and random noise can be accurately separated. If the selected threshold is too small, it may result in residual noise, making the denoising incomplete. Similarly, if the threshold is too large, while noise is effectively suppressed, some useful signal may also be damaged. After investigation, we introduce the Sigmoid function as our threshold function. The Sigmoid function can be expressed as,
        \begin{equation}
            \delta (t) = \frac{1}{{1 + {e^{ - t}}}}.
        \end{equation} 
        As shown in Fig. 6(a), when $t > 0$, $\delta (t)$ rapidly approaches 1 as $t$ increases. Conversely, when $t < 0$, $\delta (t)$ rapidly approaches 0 as $t$ decreases. Before constructing the Sigmoid threshold function, it is necessary to normalize it, placing it within the desired interval, thus obtaining a function that can be truncated (Fig. 6(b)). The truncation position is determined by the threshold $T$. The normalized Sigmoid function can be expressed as,
        \begin{equation}
            \delta (t) = \frac{1}{{1 + {e^{\frac{{-10(|t| - T)}}{T}}}}}.
        \end{equation} 
        Subsequently, we multiply eq(27) by the wave coefficient ${\cal C}$ to obtain the final Sigmoid threshold function, that is,
        \begin{equation}
            \delta ({\bm{\mathcal{C}}}) = \frac{1}{{1 + {e^{\frac{{-10(|{\bm{\mathcal{C}}}| - T)}}{T}}}}}{\bm{\mathcal{C}}}.
        \end{equation}         
        \textcolor{black}{To validate the effectiveness of curvelet denoising, we adopt the same public dataset from the first deep learning-based Massive MIMO CSI feedback work \cite{art_3}, in Fig. 7, we compare the grayscale images of the reconstructed channel at the BS when there is feedback noise, i.e., when eq(5) holds. Fig. 7(a) shows the original channel image, Fig. 7(b) shows the channel image after adding noise, and Fig. 7(c) shows the channel image after denoising with curvelets. It can be observed that Sigmoid function effectively separate the signal and noise.}

        
       \begin{figure}[t]  
            
           \centering
           \includegraphics[width=8cm]{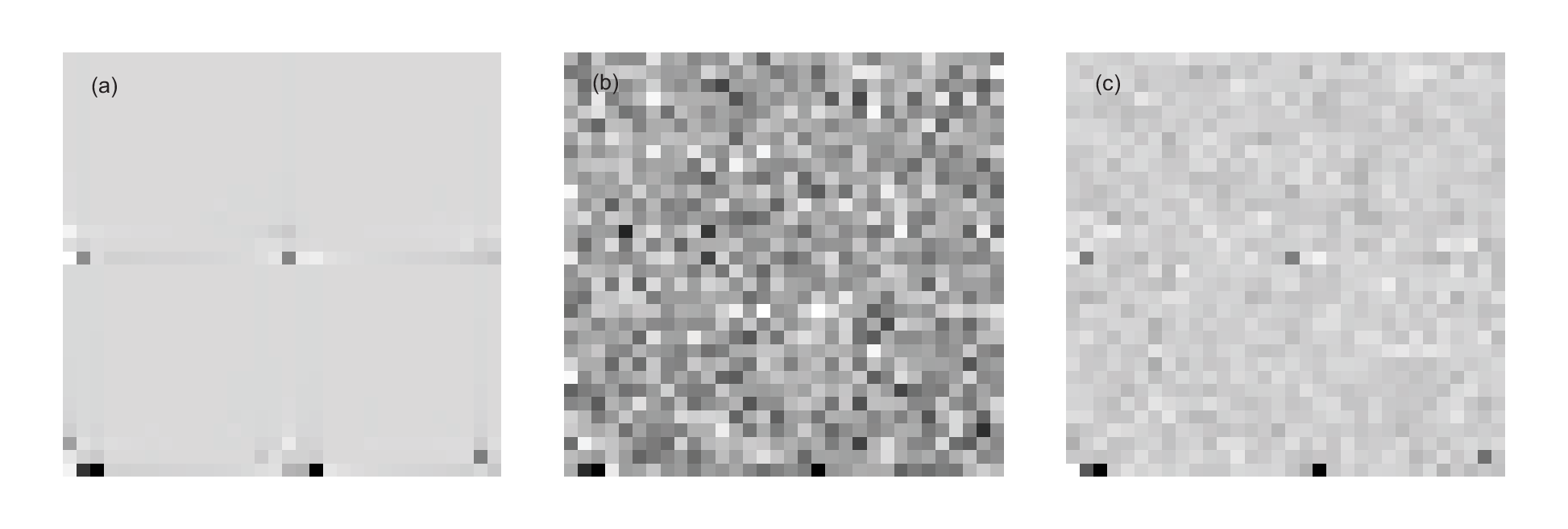}
           \captionsetup{justification=centering}
           \caption{\textcolor{black}{(a) Original channel ; (b) The noisy channel (SNR = 10 dB); (c) The denoised channel using curvelets.}}
           
      \end{figure}

    \subsection{Training of De-SwinCANet}
		To train the proposed De-SwinCANet, we employ an end-to-end approach. To minimize the difference between the original ${{{\bf H}}}$ and the reconstructed ${{{\bf{\hat H}}}}$, the loss function we use for training is the mean squared error (MSE), which can be computed as follows:
        \begin{equation}
            \textcolor{black}{{\cal L}({{\bf{H}}},{{{\bf{\hat H}}}}) =\mathbb{E} [||{{\bf{H}}} - {{{\bf{\hat H}}}}||_2^2].}
	  \end{equation}
	
        The training process of the whole system is given in Algorithm 2. The network is trained over epochs, which will converge with the training process.

\begin{algorithm}[h]
    \caption{Training Process of CSI Feedback Based on De-SwinCANet Network.}
    {\bf Input:} Feedback of batch size $n$, CSI $\bf{H}$.
    \begin{algorithmic}[1]
        \While{the training stop criterion is not met}
            \Statex {\bf Transmitter:}
            \State Curvelet transform: ${\rm{Curvelet}}_J({\bf{H}}) \rightarrow {\bm{\mathcal{C}}}$.
            \State SwinCANet encoder: ${\cal F}_\varepsilon({\bm{\mathcal{C}}}; \Theta_\varepsilon) \rightarrow \bf{g}$.
            \State Transmit $\bf{g}$ over the channel $\rightarrow {\bf{\tilde g}}$.
            \Statex {\bf Receiver:}
            \State SwinCANet decoder: ${\cal F}_\chi({\bf{\tilde g}}; \Theta_\chi) \rightarrow \tilde{\bm{\mathcal{C}}}$.
            \State Sigmoid threshold function: $\delta(\tilde{\bm{\mathcal{C}}}) \rightarrow \hat{\bm{\mathcal{C}}}$.
            \State Inverse curvelet transform: ${\rm{Inv\_Curvelet}}_J(\hat{\bm{\mathcal{C}}}) \rightarrow {\bf{\hat H}}_a$.
            \State Calculate the gradient based on (27).
            \State Train the network by gradient descent.
        \EndWhile
    \end{algorithmic}
    {\bf Output:} The whole network \\
    ${\cal F}_\varepsilon(\cdot), {\cal F}_\chi(\cdot)$.
\end{algorithm}

        \begin{figure*}[t]  
            
            \centering
            \includegraphics[width=16cm]{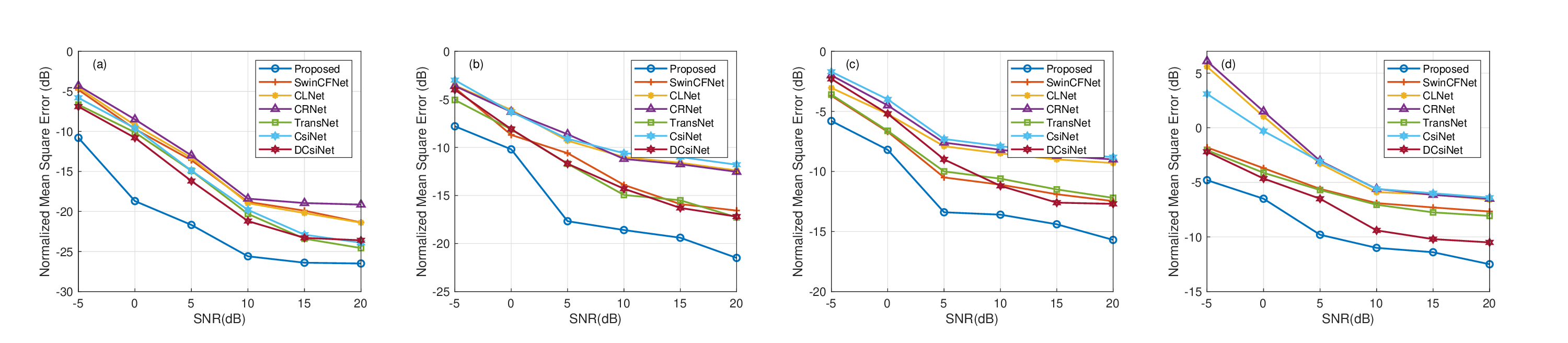}
            \captionsetup{justification=centering}
            \caption{NMSE Performance Comparison of Various CSI Reconstruction Methods with the Proposed Model under Channel Noise at Different Compression Ratios: (a) CR = 1/4, (b) CR = 1/8, (c) CR = 1/16, and (d) CR = 1/32.}
           
        \end{figure*}

        \begin{figure*}[t]  
            
            \centering
            \includegraphics[width=16cm]{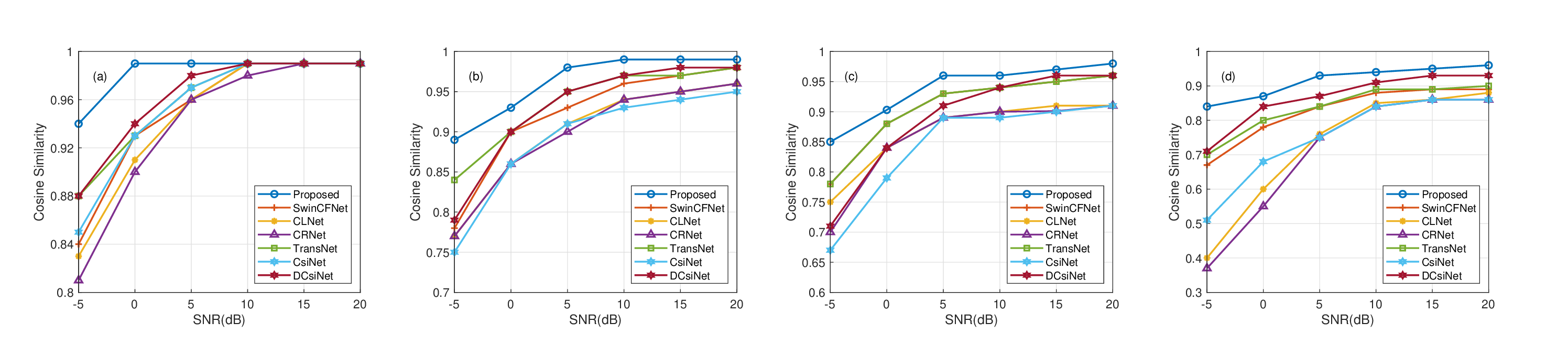}
            \captionsetup{justification=centering}
            \caption{Cosine Similarity Performance Comparison of Various CSI Reconstruction Methods with the Proposed Model under Channel Noise at Different Compression Ratios: (a) CR = 1/4, (b) CR = 1/8, (c) CR = 1/16, and (d) CR = 1/32.}
           
        \end{figure*}
        
    \section{Simulation Results}
    
	 \subsection{Simulation Settings}
     
        \emph{1) Data Generation}: In our simulation, \textcolor{black}{to more realistically evaluate the performance of the proposed CSI feedback denoising algorithm, we regenerated a dataset based on the COST2100 channel model \cite{art_29}. Existing public datasets are limited in scale and scenarios, making them insufficient to fully reflect the challenges of high CSI dimensions and large feedback overheads in large-scale antenna systems. To address this, we constructed an indoor picocellular scenario operating at 5.3 GHz,} where the system utilized a uniform linear array (ULA) with 256 antennas and 1024 subcarriers, placed at the center of a $20m \times 20m$ square area. The UE was randomly distributed within this square area, and other parameters followed the default settings in \cite{art_9}. The generated dataset contains a total of 100,000 samples, with 50,000 for training, 30,000 for validation, and 20,000 for testing. The generated CSI matrix was transformed into the angular-delay domain using 2D-DFT, and we selected the initial 32 rows, i.e., ${N_a} = 32$, converting the channel matrix into a $32 \times 256$ complex matrix in the angle-delay domain.
        
        \emph{2)Setting of training parameters and Evaluation Metric}: In our experiments, unless otherwise specified, the number of scales for the curvelet transform is set to $\lceil \log_2(\min(N_a, N_t)) - 3 \rceil$. For both the indoor and outdoor scenarios, the window size of the proposed method is set to 7 and the patch size is set to ${\rm{3}} \times {\rm{3}}$. For all experiments, the number of attention heads in each layer is set to [4, 8], and the expansion factor of each MLP layer is set to 4. The number of Swin Transformer blocks is configured as $[N_1, N_2]$ = [4, 16]. 
        The batch size and number of epoch are set to 200 and 1000, respectively. We used the Adam optimizer with a learning rate of 0.00005 and employed MSE as the loss function to improve training accuracy.
        
        In this study, the channel estimation performance is evaluated using the normalized mean squared error (NMSE),
        \begin{equation}
            {\rm{NMSE}} \buildrel \Delta \over = \mathbb{E}\left[ {\frac{{\left\| {{\bf{H}} - {\bf{\hat H}}} \right\|_2^2}}{{\left\| {\bf{H}} \right\|_2^2}}} \right].
        \end{equation}

        Cosine similarity in the spatial-frequency domain is also employed as a metric to assess and compare the effectiveness of different approaches, which is given by,
        \begin{equation}
            \rho  = \mathbb{E} \left\{ {\frac{1}{{{{\tilde N}_c}}}\sum\limits_{n = 1}^{{{\tilde N}_c}} {\frac{{|{\bf{\hat{\tilde{h}}}}_n^H{{\widetilde {\bf{h}}}_n}|{\rm{ }}}}{{||{\bf{\hat{\tilde{h}}}}_n^H|{|_2}||{{\widetilde {\bf{h}}}_n}|{|_2}}}} } \right\},
        \end{equation}
        where ${\tilde{h}}_n$ and $\hat{\tilde{h}}_n$ denote the original and reconstructed channel vector of the $n$-th subcarrier, respectively.

        \emph{3)Benchmarks}: For the baseline, we compare the proposed system with the benchmarks as follows:
        \begin{itemize} 
            \item CsiNet \cite{art_3} integrates deep learning for CSI feedback, compressing downlink CSI via an encoder for transmission.
            \item CRNet \cite{art_8} adopts a multi-resolution architecture within its network design and highlights the significance of an effective training scheme
            \item CLNet \cite{art_9} treats CSI as a physically meaningful complex-valued signal
            \item SwinCFNet \cite{art_15} based on the Swin Transformer 
            \item TransNet \cite{art_14} adopts a dual-layer Transformer architecture 
            \item DCsiNet \cite{art_22} is a deep learning model that incorporates a decoder for simultaneous CSI decompression and denoising, enhancing feedback accuracy in noise-contaminated environments.
        \end{itemize}

        \textcolor{black}{Note that all benchmark methods were retrained by us on the newly generated dataset (256 antennas $\times$ 1024 subcarriers) to ensure fair and comparable experimental evaluation.}


	\subsection{Performance Comparison}

	\begin{figure}[t]  
            
            \centering
            \includegraphics[width=8cm]{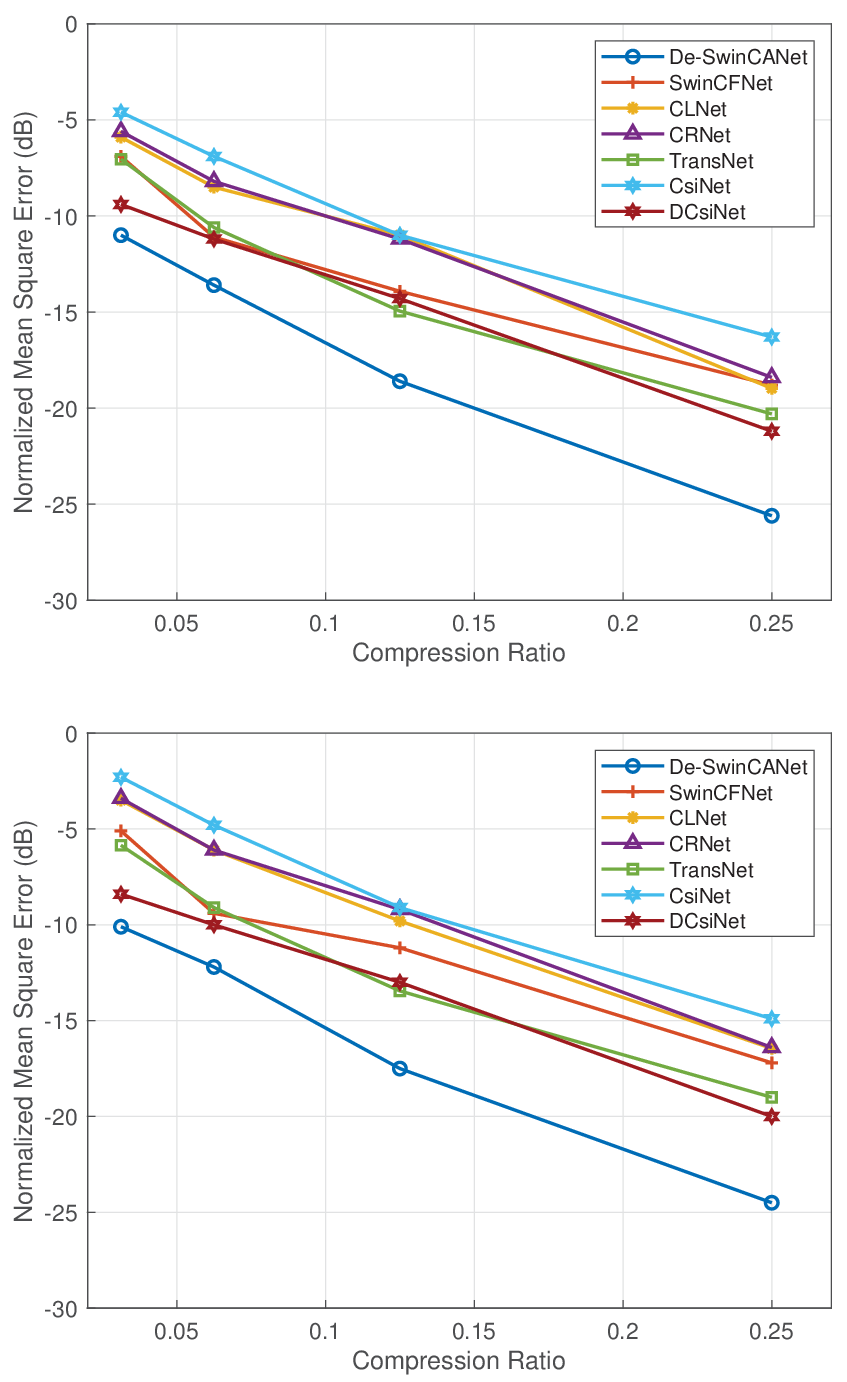}
            \captionsetup{justification=centering}
            \caption{\textcolor{black}{NMSE comparison of different models under varying CRs at an SNR = 10dB: (a) AWGN channels and (b) Rayleigh fading channels.}}
           
        \end{figure}       
        
	\begin{figure}[t]  
            
            \centering
            \includegraphics[width=8cm]{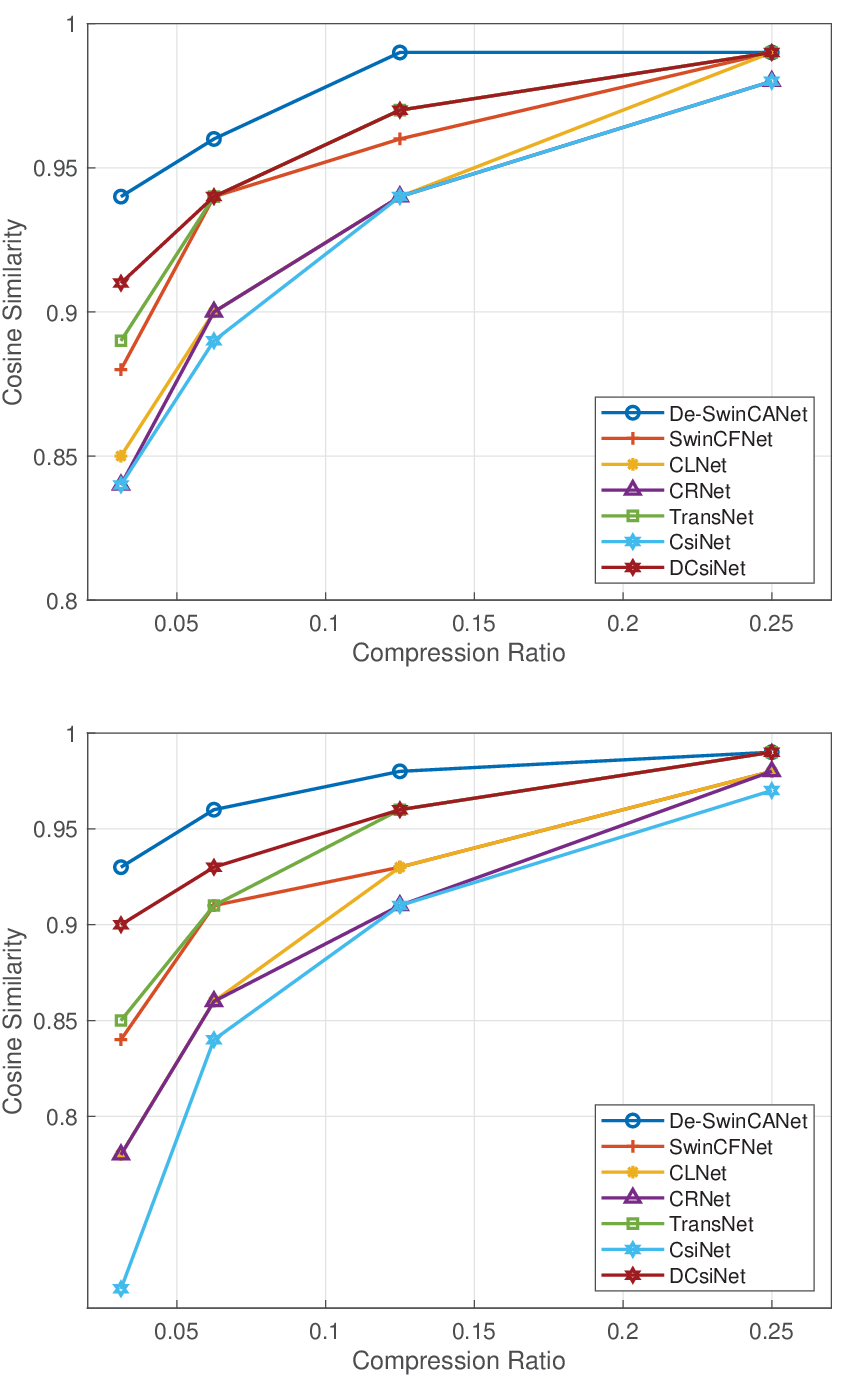}
            \captionsetup{justification=centering}
            \caption{\textcolor{black}{Cosine similarity comparison of different models under varying CRs at an SNR = 10dB: (a) AWGN channels and (b) Rayleigh fading channels.}}
           
        \end{figure}
        
        Fig. 8 compares the NMSE performance of various methods under different SNR and compression ratio (CR) conditions. As expected, performance improves with increasing CR and SNR. The proposed De-SwinCANet consistently outperforms existing CSI reconstruction methods and the denoising-enabled DCsiNet across all settings, exhibiting superior robustness even under high compression or strong noise. \textcolor{black}{The main reason is that the proposed method decomposes the CSI into low-frequency and high-frequency components by curvelet transform \cite{art_23} and encodes and decodes the low-frequency and high-frequency components by using Swin Transform and channel-wise attention module.} Such separation enhances feature extraction efficiency, integrates global semantics with local details, reduces information redundancy, and optimizes computational resource allocation. Moreover, the incorporation of a thresholding function mitigates noise in the feedback loop, further improving reconstruction accuracy. Notably, at SNR = 10 dB, the proposed method surpasses all competing methods operating at SNR = 20 dB across all CRs, underscoring its exceptional robustness and generalization capability.

        Meanwhile, we further present the comparison of $\rho$ values under different SNR conditions and CR across various methods in Fig. 9. Consistent with the NMSE performance under the same conditions in Fig. 8, the proposed method also shows a gradual improvement in $\rho$ values as the compression ratio and SNR increase. In contrast, thanks to the introduction of the curvelet transform, our approach employs the Swin Transformer and channel-wise attention mechanism to model the low- and high-frequency components of CSI, respectively, significantly enhancing feature extraction accuracy. Additionally, the use of a thresholding function effectively suppresses noise interference, enabling our method to outperform other approaches across all compression ratios. Notably, in challenging scenarios such as high compression or severe noise conditions, the proposed method maintains high reconstruction accuracy, demonstrating superior robustness against compression and noise.

         To evaluate robustness, Fig. 10 compares the NMSE performance of the proposed De-SwinCANet against existing CSI feedback networks and the denoising model DCsiNet under AWGN and Rayleigh fading channels at SNR = 10 dB across compression ratios of 1/32, 1/16, 1/8, and 1/4. As expected, NMSE improves with increasing CR for all methods. The proposed method consistently outperforms others, particularly at lower CRs, achieving NMSE below −20 dB at CR = 1/4. This advantage is attributed to curvelet transform-based decomposition of CSI into low- and high-frequency components, processed separately via Swin Transformer and channel-wise attention, alongside a thresholding function for noise suppression. Under fading channels, all methods exhibit performance degradation relative to AWGN; however, the proposed method demonstrates superior stability, with only approximately 1 dB increase in NMSE, validating its robustness. The separate processing strategy enhances generalization and resilience in complex CSI environments.

\begin{table*}[htbp]
\centering
\caption{\textcolor{black}{Performance and Complexity Comparison of Methods under Antenna Configurations (CR=1/4, SNR=10dB, AWGN)}}
\label{tab:comparison}
\begin{tabular}{c|cc|cc|cc|cc|cc}
\hline
\multirow{2}{*}{Antennas} & \multicolumn{2}{c|}{Proposed} & \multicolumn{2}{c|}{SwinCFNet} & \multicolumn{2}{c|}{CLNet} & \multicolumn{2}{c|}{TransNet} & \multicolumn{2}{c}{DsiNet} \\
\cline{2-11}
 & NMSE & FLOPS & NMSE & FLOPS & NMSE & FLOPS & NMSE & FLOPS & NMSE & FLOPS \\
\hline
32  & -28.1 & 80.671M  & -24.4 & 624.12M  & -23.5 & 4.05M   & -26.2 & 35.75M  & -25.7 & 292.864M \\
64  & -27.7 & 152.742M & -23.3 & 1.287G   & -22.1 & 13.271M & -24.9 & 75.694M & -24.9 & 745.268M \\
128 & -27.1 & 319.03M  & -21.2 & 2.540G   & -20.6 & 43.319M & -22.8 & 168.165M & -24.1 & 1.685G \\
256 & -25.6 & 511.576M & -18.8 & 5.045G   & -18.97 & 153.748M & -20.3 & 403.44M & -21.2 & 4.032G \\
\hline
\end{tabular}
\end{table*}

        Fig. 11 presents a comparison of the NMSE and $\rho$ performance at SNR = 10 dB of the proposed De-SwinCANet with the associated CSI feedback networks under different compression ratios over AWGN channel and fading channel for CSI feedback. The compression ratios are set to 1/32, 1/16, 1/8, and 1/4. From the figure, we can observe that the $\rho$ values for all methods increases as the compression ratio increases and our method outperform all other methods, especially at lower compression ratios. And we can get the same conclusion as Fig. 11, all other methods show performance degradation under fading channel with respect to AWGN channel, while the proposed method remains basically stable, which proves the robustness of the proposed method.

	\textcolor{black}{To evaluate the scalability of the proposed method with respect to antenna array size, comparative experiments were conducted under AWGN channels (CR=1/4, SNR=10dB) across various antenna configurations, with results summarized in Table I. The proposed method exhibits favorable scalability in both computational complexity and reconstruction performance. In terms of complexity, its FLOPs increase approximately linearly with the number of antennas. While SwinCFNet and DsiNet also display linear growth, their absolute computational costs remain substantially higher. TransNet incurs moderate complexity but suffers from more rapid performance degradation as antenna count grows. Regarding reconstruction quality, the proposed method achieves the best or near-best NMSE across all configurations, with performance degradation notably smaller than that of competing approaches such as SwinCFNet and TransNet. These results demonstrate that the proposed method strikes an effective balance between performance and complexity, achieving superior reconstruction accuracy with moderate computational overhead, rendering it well-suited for large-scale MIMO systems with stringent efficiency requirements.}

    \subsection{\textcolor{black}{Performance Analysis of Quantization and Entropy Coding}}

\enlargethispage{2\baselineskip}

    \textcolor{black}{In practical communication systems, the compressed codeword $\mathbf{g}$ needs to be quantized before being transmitted over digital feedback channels. To verify the applicability of the proposed method in practical bit-level transmission scenarios, this section introduces an entropy bottleneck layer \cite{art_40, art_41} to analyze the impact of quantization and entropy coding on system performance. We adopt the training configuration outlined in Section IV-A throughout our experiments. The compression ratio is fixed at 1/4, corresponding to an encoder output dimension of $k=4096$. The training is performed under AWGN channel conditions with a SNR of 10 dB. To characterize the rate-distortion performance, we train multiple networks with the regularization parameter $\lambda$ logarithmically distributed over the interval $[10^{-1}, 10^5]$. The quality of channel reconstruction is quantified by the NMSE defined in Eq. (18). The principles of quantization and entropy coding are described in detail.\protect\footnote{https://github.com/MenliTao/Principles-of-Quantization-and-Entropy-Coding/blob/main/Supplementary\%20Materials.pdf}}
    

    	\begin{figure}[t]  
            
            \centering
            \includegraphics[width=8cm]{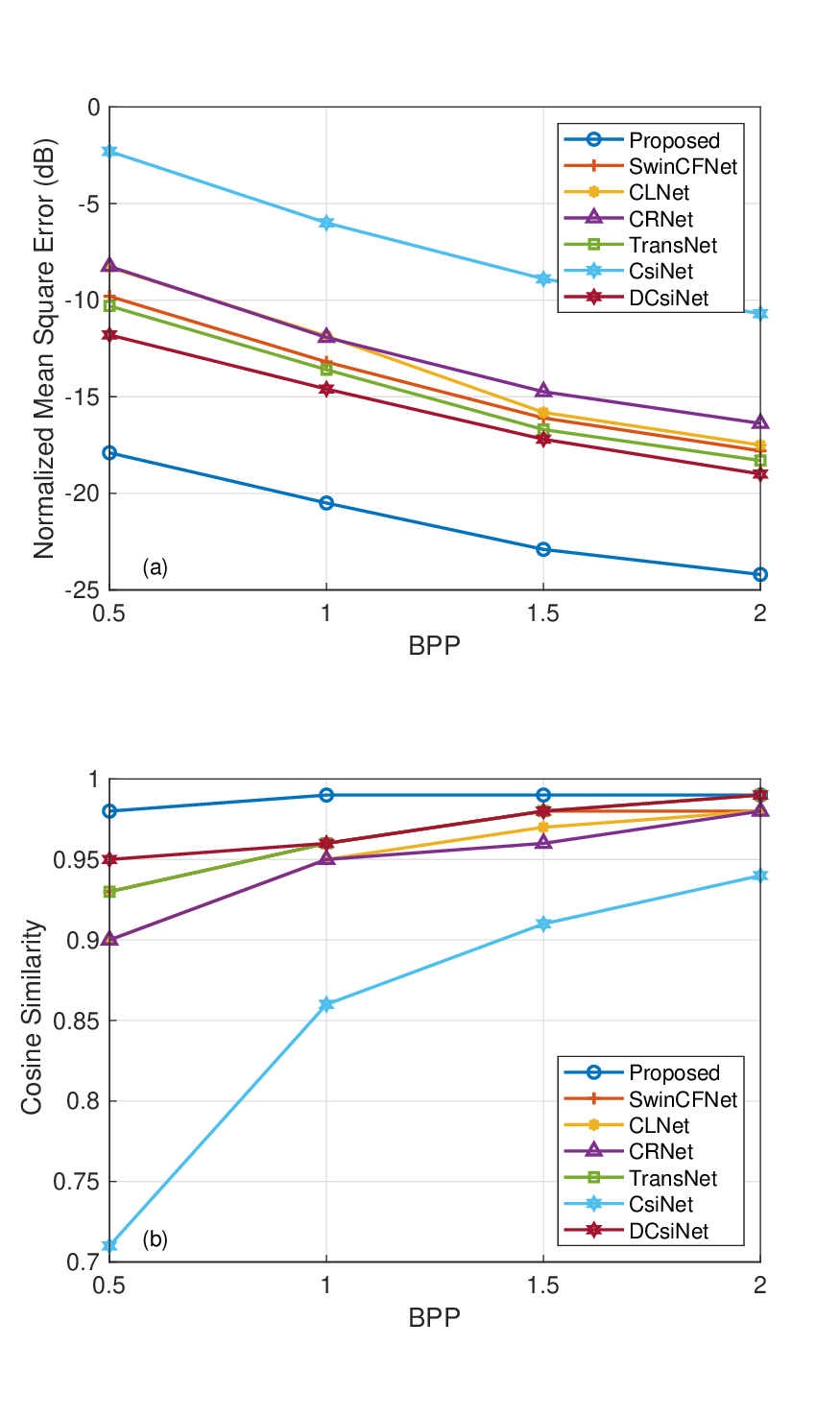}
            \captionsetup{justification=centering}
            \caption{\textcolor{black}{\textcolor{black}{Rate-distortion performance comparison of different CSI feedback methods at an SNR = 10dB: (a) NMSE; (b) Cosine similarity.}}}
           
        \end{figure}

\begin{table*}[h]
    \centering
    \renewcommand{\arraystretch}{1.3}
    \caption{Comparison of NMSE and $\rho$ in ablation experiments under different compression ratios at SNR=20dB.}
    \begin{tabular*}{14cm}{@{\extracolsep{\fill}}|c|cc|cc|cc|cc|}
        \hline
        CR & \multicolumn{2}{c|}{cr=1/4} & \multicolumn{2}{c|}{cr=1/8} & \multicolumn{2}{c|}{cr=1/16} & \multicolumn{2}{c|}{cr=1/32} \\
        \hline
        Performance & NMSE & $\rho$ & NMSE & $\rho$ & NMSE & $\rho$ & NMSE & $\rho$ \\
        \hline
        De-SwinCANet & $\bm{-25.9}$ & $\bm{0.99}$ & $\bm{-21.5}$ & $\bm{0.99}$ & $\bm{-15.6}$ & $\bm{0.98}$ & $\bm{-12.5}$ & $\bm{0.96}$ \\
        Without curvelet transform & $-23.7$ & 0.99 & $-18.9$ & 0.99 & $-13.6$ & 0.96 & $-10.1$ & 0.93 \\
        Without CAB & $-22.1$ & 0.99 & $-18.0$ & 0.98 & $-13.1$ & 0.96 & $-9.6$ & 0.93 \\
        Without threshold function & $-24.5$ & 0.99 & $-19.9$ & 0.99 & $-14.5$ & 0.97 & $-11.0$ & 0.94 \\
        \hline
    \end{tabular*}
\end{table*}

        \textcolor{black}{Fig. 12 shows the rate-distortion performance comparison of different CSI feedback methods. As shown in Fig. 12, we compared the rate-distortion performance of the proposed method with six benchmark methods across a  bits per pixel (BPP) range from 0.5 to 2. In terms of the NMSE metric, the proposed method outperforms all comparison methods throughout the entire BPP range. Specifically, at BPP=0.5, the proposed method achieves an NMSE of approximately -18 dB, which is about 8 dB lower than SwinCFNet and nearly 12 dB lower than CsiNet. When BPP increases to 2, the proposed method reaches approximately -24 dB, maintaining a performance gap of about 5 dB compared to SwinCFNet. Notably, methods such as CsiNet, TransNet, and CRNet exhibit relatively slow NMSE reduction as BPP increases, whereas the proposed method demonstrates a steeper downward trend, indicating its ability to utilize additional bits more efficiently. The cosine similarity results further validate the above findings. The proposed method maintains high similarity across all BPP values. Although the similarity of all methods tends to converge at BPP=2, the performance gap at lower BPP values highlights the superior compression efficiency of the proposed method. In summary, the proposed method demonstrates the best rate-distortion trade-off performance among all evaluated methods.}

\begin{table}[h]
    \centering
    \caption{\textbf{\textcolor{black}{Complexity comparison between series of CSI feedback network and the proposed De-SwinCANet}}}
    \renewcommand{\arraystretch}{1.5}
    \setlength{\tabcolsep}{2.0pt}
    \begin{tabular}{|>{\centering\arraybackslash}m{2.0cm}|>{\centering\arraybackslash}m{2.0cm}|>{\centering\arraybackslash}m{2.0cm}|>{\centering\arraybackslash}m{2.0cm}|}
        \hline
        \textbf{Method} & \textbf{FLOPs (G)} & \textbf{Params (M)} & \textbf{NMSE (dB)} \\
        \hline
        SwinCFNet & 5.045 & 7.541  & -18.8  \\
        \hline
        TransNet  & 0.403 & 2.629  & -20.3  \\
        \hline
        CLNet     & 0.154 & 134.240 & -18.97 \\
        \hline
        CRNet     & 0.161 & 134.241 & -18.4  \\
        \hline
        DCsiNet   & 4.032 & 789.213 & -21.2  \\
        \hline
        CsiNet    & 0.096 & 67.129  & -16.3  \\
        \hline
        Proposed  & 0.512 & 2.107   & -25.6  \\
        \hline
    \end{tabular}
    \label{tab:complexity_comparison}
\end{table}

    \subsection{Ablation Experiment}

	To verify the effectiveness and necessity of each component in the model, we conducted three groups of ablation experiments in this section: (1) Ablation of curvelet transform (Without curvelet transform): directly applying the proposed SwinCANet to the CSI feedback task; (2) Ablation of channel-wise attention module (Without CAM): using Swin Transformer to process the low-frequency and high-frequency information decomposed by curvelet transform for feedback; (3) Ablation of threshold function (Without threshold function): removing the use of threshold function, where the curvelet coefficients recovered by SwinCANet decoder directly undergo inverse curvelet transform to obtain the reconstructed CSI matrix.

    Table II presents ablation results for curvelet transform, channel-wise attention module, and threshold function under different CRs. Ablating each component causes performance degradation in both NMSE and $\rho$, with NMSE decline being more pronounced. \textcolor{black}{The ablation of the curvelet transform results in the most significant NMSE increase because it plays a fundamental role in frequency-domain decomposition, effectively separating low-frequency information (main structural features) from high-frequency information (edge and texture details) for differentiated processing. Without the curvelet transform, the model is forced to process all frequency components uniformly, substantially reducing its ability to capture the complex frequency characteristics of CSI. This result demonstrates that the multi-scale and multi-directional decomposition capability provided by the curvelet transform cannot be effectively replaced by end-to-end neural network learning, as the curvelet transform provides a mathematically optimized framework whose anisotropic characteristics are highly matched with massive MIMO channel properties, while learning such complex structures from scratch requires enormous model capacity and training data.} After removing the channel-wise attention module, using only Swin Transformer to process the curvelet-transformed information also leads to performance degradation. This is because high-frequency information has characteristics of strong locality and dramatic variations, requiring sophisticated local feature capture capabilities. The channel-wise attention module is specifically designed to capture local details in high-frequency information, while Swin Transformer's window attention mechanism is mainly suitable for processing low-frequency information with global structural characteristics and cannot adequately capture local detail features in high-frequency components. Although removing the threshold function results in relatively mild performance degradation, it still causes performance loss. The threshold function plays an important denoising role in the curvelet domain, adaptively suppressing noise coefficients and redundant information while preserving the most important coefficients for signal reconstruction. Without the threshold function, coefficients containing noise accumulate errors during the reconstruction process, reducing the reconstruction quality of the CSI matrix. 
    
    In summary, ablation studies validate the effectiveness and necessity of each model component. The curvelet transform, channel-wise attention module, and threshold function fulfill irreplaceable roles in frequency-domain decomposition, local feature extraction, and denoising, respectively. Their synergy underpins CSI feedback performance, and removing any component degrades results, confirming the rationality and efficiency of the proposed architecture.

    \subsection{\textcolor{black}{Complexity Analysis}}

    \textcolor{black}{Table III compares the computational complexity of the proposed method with existing CSI feedback approaches at CR = 1/4. In terms of parameter count, the proposed method requires only 2.107M parameters, significantly fewer than CNN-based methods such as CLNet, CRNet, and CsiNet, and also more efficient than TransNet (2.629M), demonstrating a clear lightweight advantage suitable for resource-constrained mobile devices. From the perspective of floating-point operations (FLOPs), the computational load of the proposed method is 0.512G. Although slightly higher than CLNet, CRNet, and TransNet, it is substantially lower than Transformer-based architectures such as SwinCFNet and DCsiNet. Notably, FLOPs are not a practical bottleneck in CSI feedback systems \cite{art_14}; the core objective remains reconstruction accuracy. The proposed method achieves superior NMSE performance while maintaining low parameter overhead, validating the effectiveness of its architectural design. Overall, it strikes a favorable balance between computational efficiency and high-precision reconstruction.}

		\section{Conclusion}

        In this paper, we proposed a channel feedback denoising model named De-SwinCANet for FDD massive MIMO systems. The proposed De-SwinCANet was based on the curvelet transform to separate the low-frequency and multi-scale high-frequency information of CSI, which can reduce the learning difficulties of CSI compression. Then, we processed this information by using SwinTransformer and the designed channel-wise attention module for low-frequency and high-frequency components, respectively. Furthermore, we fused the processed features with another SwinTransformer to remove redundancy between low-frequency and high-frequency components. At the receiver, we designed the denoising threshold function to remove the channel noise and improve the accuracy of the received CSI. The numerical results shown that the proposed De-SwinCANet achieves the state-of-the-art performance in terms of NMSE with fewer transmitted symbols and is more robust in low SNR regions.

\balance
\bibliographystyle{IEEEtran}

\bibliography{IEEEabrv,reference1}

\end{document}